\documentclass[11pt]{article}

\usepackage{epsfig}
\usepackage{subfig}
\usepackage{graphicx}
\usepackage{epstopdf}
\usepackage{amsfonts}
\usepackage{amssymb}
\usepackage{amsthm}
\usepackage{amsmath}
\usepackage{multirow}
\usepackage{color}
\usepackage{cite}
 \usepackage{makecell}
\def\beq{\begin{equation}}
\def\eeq{\end{equation}}
\def\bea{\begin{eqnarray}}
\def\eea{\end{eqnarray}}
\newcommand{\beqs}{\begin{subequations}}
\newcommand{\eeqs}{\end{subequations}}

\newcommand{\cref}[1]{Ref.~\cite{#1}}

\newcommand{\hh}{{\ensuremath{I{\kern-2.6pt h}}}}
\newcommand{\bhh}{{\ensuremath{\bar{I{\kern-2.6pt h}}}}}

\renewcommand{\figurename}{Fig.}

\newcommand{\Z}{\mathbb{Z}}

\usepackage[colorlinks=true,allcolors=blue]{hyperref}

\begin{document}

\begin{titlepage}
	

\begin{center}
{\Large {\bf Topological structures, dark matter \\ and gravitational waves in $E_6$}
}
\\[12mm]
Rinku Maji,$^{1}$
Qaisar Shafi,$^{2}$
Amit Tiwari$^{2}$
\end{center}
\vspace*{0.50cm}
	\centerline{$^{1}$ \it
		Cosmology, Gravity and Astroparticle Physics Group, }
  \centerline{\it  Center for Theoretical Physics of the Universe,}
		\centerline{\it  Institute for Basic Science, Daejeon 34126, Republic of Korea}
	\vspace*{0.2cm}
	\centerline{$^{2}$ \it
		Bartol Research Institute, Department of Physics and 
		Astronomy,}
	\centerline{\it
		 University of Delaware, Newark, DE 19716, USA}
	\vspace*{1.20cm}

\begin{abstract}
We discuss the appearance of topological structures from the spontaneous breaking of $E_6$ to the Standard Model via its maximal subgroup $SO(10) \times U(1)_\psi$. They include dumbbells, metastable strings, as well as domain walls bounded by necklaces. We provide a novel scenario for producing metastable strings based on the symmetry breaking $U(1)_\psi \longrightarrow \Z_8 \longrightarrow \Z_4$. The metastable string arises from the merger of $\Z_8$ strings that bound a domain wall. 
An unbroken gauge $\Z_2$ symmetry from $SO(10)$ breaking yields viable stable dark matter candidates as well as topologically stable strings. We discuss the gravitational wave emission from two varieties of cosmic strings, namely the superheavy metastable ones and the intermediate scale topologically stable cosmic strings.
\end{abstract}

\end{titlepage}
\section{Introduction}
In $SO(10)$, more precisely $Spin (10)$ grand unification \cite{Georgi:1974my,Fritzsch:1974nn}, there are two important discrete gauge symmetries.
The first one, a $\Z_2$ gauge symmetry, is a subgroup of the center $\Z_4$ of $SO(10)$ which, if left unbroken, yields topologically stable cosmic strings \cite{Kibble:1982ae}. In a supersymmetric framework this $\Z_2$ is precisely ``matter" parity, which implies the stability of the lightest supersymmetric particle  and thus its appearance as a compelling dark matter candidate. Even in the absence of supersymmetry this $\Z_2$ symmetry can guarantee the existence of dark matter particle depending on the particle content of the model. For instance, by supplementing $SO(10)$ with a $U(1)$ axion symmetry and requiring absence of the axion domain wall problem, one is led to introduce additional matter fields in the model that yield a WIMP-like dark matter in addition to axions \cite{Lazarides:1982tw,Holman:1982tb, Lazarides:2020frf}. For other Refs. on dark matter in SO(10) model see, \cite{Kadastik:2009dj,Mambrini:2015vna,Boucenna:2015sdg,Ferrari:2018rey,Bhattacharyya:2021lgr,Lazarides:2022spe,Lazarides:2022ezc,Sahu:2022rwq,Bhattacharyya:2022trp,Okada:2022yvq}. 

The second discrete gauge symmetry in $SO(10)$ which plays an important role in cosmology and particle physics is $C$-parity \cite{Kibble:1982dd,Lazarides:1985my}, also referred to as $D$-parity \cite{Chang:1983fu}. In contrast to the above discrete $\Z_2$ symmetry, the $C$-parity is necessarily spontaneously broken, and depending on the $SO(10)$ symmetry breaking pattern, one predicts the appearance of composite topological structures known as ``walls bounded by strings" (WBS) \cite{Kibble:1982dd}. Remarkably, WBS have recently been discovered in superfluid $^{3}$He-B phase transitions \cite{Makinen:2018ltj}. These composite structures, now found to be present in a wide variety of realistic unified models \cite{Dunsky:2021tih,Maji:2023fba,Eto:2023aqr, Roshan:2024qnv, Fu:2024rsm,Hamada:2024dan}, may have left traces of their presence in the early universe through the emission of gravitational waves. 

In a recent paper \cite{Lazarides:2023iim} we made a systematic study of composite topological structures that can appear in realistic $SO(10)$ models. In addition to structures such as dumbbells and WBS, we identified some more exotic objects such as ``walls bounded by necklaces" and ``starfish". Here, a necklace refers to a closed string with monopoles as beads on it \cite{Hindmarsh:1985xc,Aryal:1987sn,Kibble:2015twa}. The ``starfish" configuration consists of a higher charge monopole with flux tubes ending on five distinct lower charge antimonopoles \cite{Lazarides:2023iim}. 

Our aim in this paper is to extend the $SO(10)$ discussion started in \cite{Lazarides:2023iim} to $E_6$ grand unification \cite{Gursey:1975ki,Achiman:1978vg,Shafi:1978gg}. For some earlier work on topological structures in $E_6$, see Ref.~\cite{Lazarides:2019xai}. We focus here on the breaking of $E_6$ via its subgroup $SO(10) \times U(1)_\psi$, and the various topological structures we identify here are closely linked with the abelian component $U(1)_\psi$. In addition to a practically stable $U(1)_\psi$ string, first identified by Witten \cite{Witten:1984eb}, we highlight dumbbells, superheavy metastable strings \cite{NANOGrav:2023hvm,Buchmuller:2023aus,Antusch:2023zjk,Lazarides:2023rqf,Maji:2023fhv,Ahmed:2023rky,Afzal:2023cyp,Afzal:2023kqs,Ahmed:2023pjl,Lazarides:2023bjd,Pallis:2024mip} which can explain the recent PTA observations \cite{NANOGrav:2023gor,EPTA:2023fyk,Reardon:2023gzh,Xu:2023wog}, and domain wall bounded by a necklace. We present a novel well motivated scenario for producing metastable cosmic strings based on the symmetry breaking $E_6 \longrightarrow [SO(10) \times U(1)_\psi] / \Z_4 \longrightarrow [SO(10) \times \Z_8] / \Z_4 \longrightarrow SO(10)$. In the presence of primordial $U(1)_\psi$ monopoles this breaking pattern  leads to the appearance of a domain wall bounded by a necklace. However, if the primordial monopoles are inflated away, we can obtain metastable strings associated with flux tubes $\psi / 4$ from the merger of two $\Z_8$ strings carrying $\psi / 8$ flux each, that are initially separate but attached to the same domain wall. The quantum mechanical tunneling of $U(1)_\psi$ monopole-antimonopole pairs carrying magnetic charges $\pm \psi / 4$ makes the string metastable. Of course, if the $U(1)_\psi$ monopoles are much heavier than the breaking scale of $U(1)_\psi$, the quantum mechanical tunneling is exponentially suppressed and these strings are practically stable.

By retaining an unbroken $\Z_2$ symmetry from the $SO(10)$ subgroup of $E_6$, we also produce topologically stable strings and WIMP-like dark matter candidates. We also identify a scenario with keV scale sterile DM neutrino. We discuss the gravitational wave spectrum generated by metastable cosmic strings with $G\mu \sim 10^{-7}$ accompanied by topologically stable cosmic strings with $G\mu \sim 10^{-12} - 10^{-10}$, where $G\mu$ denotes the dimensionless string tension parameter. This will be tested by the ongoing pulsar timing array experiments as well as other ongoing and proposed experiments.

\section{$E_6$ Model and $U(1)_\psi$ monopoles}
\label{sec:sec2}
We are primarily focused on topological structures that arise from the symmetry breaking $E_6 \to SO(10) \times U(1)_\psi  / \Z_4 \to SO(10)$. There will be some overlap with the discussion carried out in Ref.~\cite{Lazarides:2019xai} which also considers the breaking of $E_6$ via the trinification subgroup $SU(3)_c \times SU(3)_L \times SU(3)_R$.
\begin{table}[htbp]
\centering
\scalebox{1.0}{
\begin{tabular}{|c|c|c|c|}

\hline
\hline
{Matter fields} & {Representations under $G_{\rm SM}$} & {$2\sqrt{6}\,Q_{\psi}$}\\

{} & {} & {}\\

\hline

{$q_i$} & {$({\bf 3, 2}, 1/6)$} & $1$\\

\hline

{$u_i^c$} & {$({\bf \bar 3, 1},-2/3)$} & $1$\\

\hline

{$d_i^c$} & {$({\bf \bar 3, 1},1/3)$} & $1$\\

\hline

{$l_i$} & {$({\bf 1, 2}, -1/2)$} & $1$\\

\hline

{$\nu_i^c$} & {$({\bf 1, 1}, 0)$} & $1$\\

\hline

{$e_i^c$} & {$({\bf 1, 1}, 1)$} & $1$\\

\hline

{$\tilde{H}_{ui}$} & {$({\bf 1, 2},1/2)$} & $-2$\\

\hline

{$\tilde{H}_{di}$} & {$({\bf 1, 2},-1/2)$} & $-2$\\

\hline

{$D_i$} & {$({\bf 3, 1},-1/3)$} & $-2$\\

\hline

{$D_i^c$} & {$({\bf \bar 3, 1},1/3)$} & $-2$\\

\hline

{$N_i$} & {$({\bf 1, 1},0)$} & $4$\\

\hline
\hline
\end{tabular}
}
\caption{Transformation properties of the $E_6$ matter fields under the SM gauge group $G_\text{SM}$ and their charges under the local symmetry $U(1)_{\psi}$, where $Q_{\psi}$ is the normalized GUT generator of $U(1)_\psi$. Family indices are denoted by the subscript $i\;(=1,2,3)$.}
\label{tab:fields}
\end{table}
 The matter content of the model consisting of three chiral 27- plets is shown in Table \ref{tab:fields} which also lists their $U(1)_\psi$ quantum numbers. The generator is normalized such that a minimal charge of unity is compatible with a period of $2 \pi$.
\begin{figure}[htbp]
\centering
      \includegraphics[width=0.4\textwidth,angle=0]{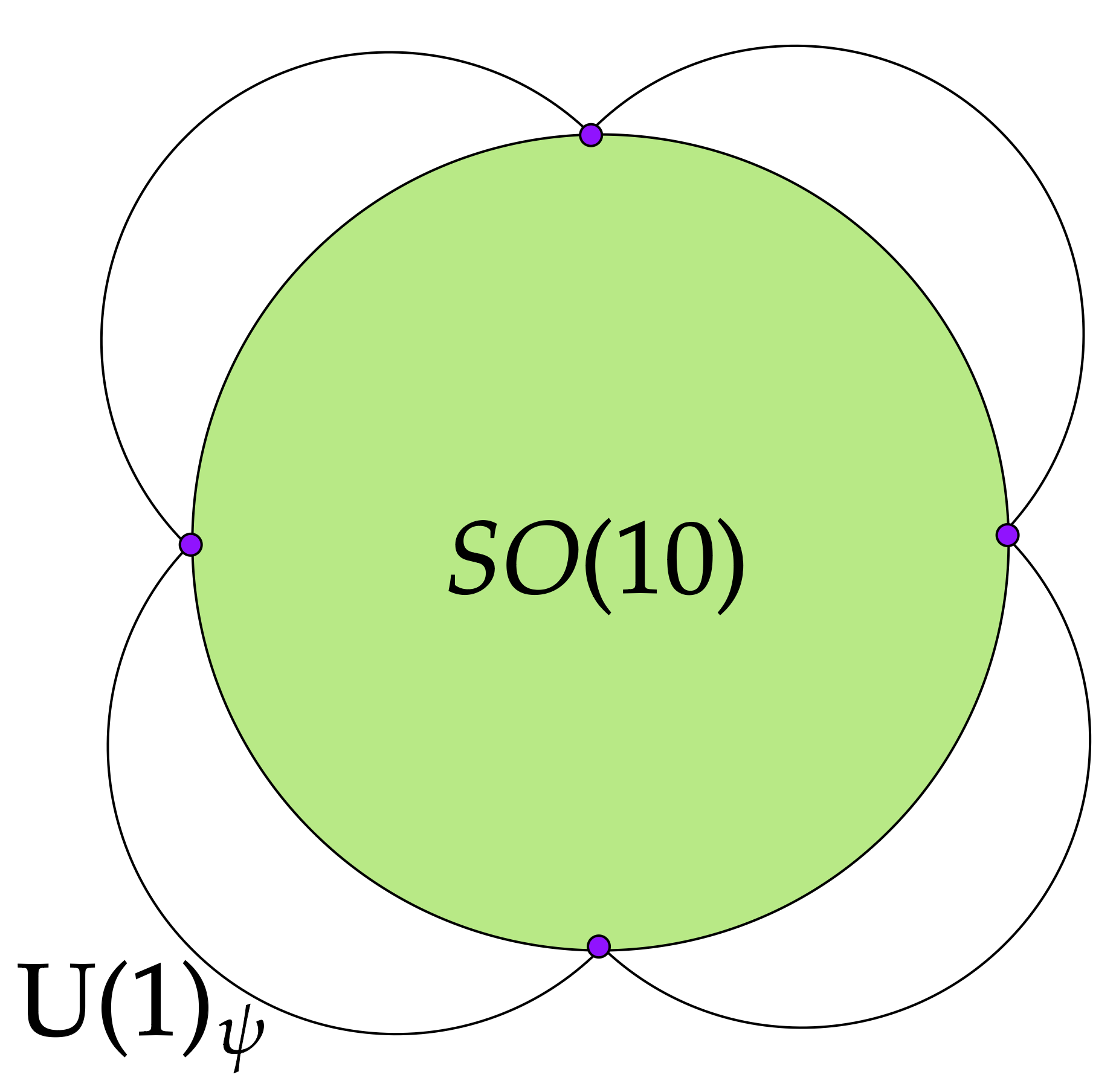}
        \caption{Embedded inside $E_6$, $U(1)_\psi$ intersects with the $SO(10)$ group in the center $\Z_4$. Minimally charged $U(1)_\psi$ monopole corresponds to a $2\pi/4$ rotation along this circle.}
  \label{fig:spin10u1}
\end{figure}

The breaking of $E_6$ to $SO(10) \times U(1)_ \psi /\Z_4$ is implemented with a Higgs 78-plet. The group $U(1)_\psi$ intersects $SO(10)$ in the center $\Z_4$ \cite{Lazarides:2019xai}, and so the minimally charged monopole produced in this breaking corresponds to $2 \pi / 4$ rotation along $U(1)_\psi$, followed by an appropriate $SO(10)$ rotation to return to the identity element, as shown in Fig.~\ref{fig:spin10u1}. This monopole carries $1/4$ of $U(1)_\psi$ magnetic flux as well as some $SO(10)$ flux.

 Next let us consider the consequence of a full $2 \pi$ rotation along $U(1)_\psi$. Clearly, this yields a monopole that carries one full unit of $U(1)_\psi$ flux (we will denote this flux simply by $\psi$). Anticipating the Higgs phase associated with $U(1)_\psi$ breaking, a novel composite topological structure with finite energy can be realized by combining a unit $\psi$ flux monopole with four minimally charged antimonopoles.
\section{Dumbbells and fourfold monopole}

\begin{figure}[h!]
 \centering
      \includegraphics[width=0.4\textwidth,angle=0]{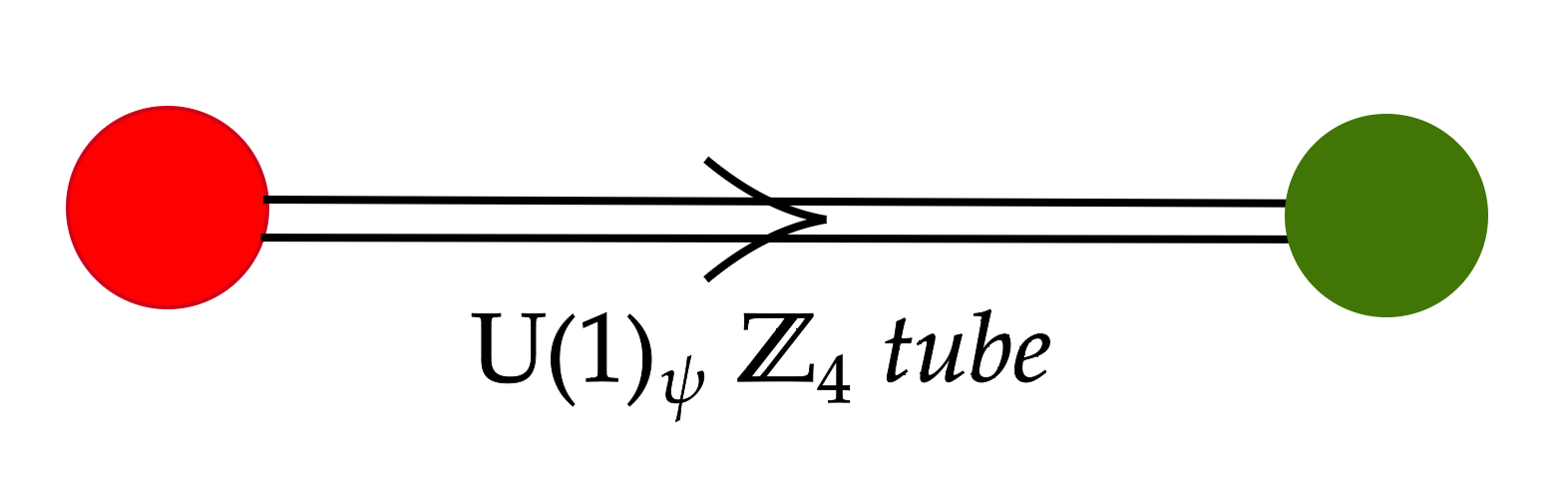}
   \caption{The second step of symmetry breaking involves giving a VEV to the $N$-type component of the Higgs 27-plet, thus breaking $U(1)_\psi$ to $\Z_4$ ($\subset SO(10)$). The flux of the minimal $U(1)_\psi$ monopole (in red) from the first step is confined in a $U(1)_\psi \; \Z_4$ tube, connected to the minimal $U(1)_\psi$ anti-monopole (in green).}
  \label{fig:dumbellonefoure6monopole}
\end{figure}

Recall that the breaking of $E_6$  to $SO(10) \times U(1)_\psi / \Z_4$ with a 78-plet vev along the singlet direction yields a minimally charged magnetic monopole that carries $\psi/4$ flux. With a non-zero vev of $N$ that carries a charge of 4 (Table \ref{tab:fields}), $U(1)_\psi$ is then broken to $\Z_4$ which is identified with the center of $SO(10)$. This breaking produces a tube with flux $\psi/4$ that confines the magnetic monopole produced in the first step. A dumbbell consisting of a $U(1)_\psi$ monopole-antimonopole pair connected by the flux tube is shown in Fig.~\ref{fig:dumbellonefoure6monopole}.

To realize the metastable string scenario that explains the PTA observations, we can assume that the two symmetry breaking scales corresponding to vevs of the 78-plet and $N$-plet are adequately close together, and the dimensionless string tension parameter $G \mu$ is of order $10^{-6}$ or so. The primordial $U(1)_\psi$  monopoles are inflated away, but they can break the $U(1)_\psi$ string through quantum mechanical tunneling, thus making it metastable. An inflationary scenario in $E_6$ can be achieved with an $E_6$ can be achieved with an $E_6$ singlet inflaton field, following the approach described in Ref.~\cite{Lazarides:2023rqf,Moursy:2024hll} based on hybrid inflation. However, a detailed treatment of inflation is beyond the scope of this paper.

Note that the fermions in the 10-plets of $SO(10)$ acquire masses through their renormalizable couplings to $N$, while the $N_i$ matter fields do so through the dimension five couplings $N_i N_j \left<N\right> \left<N\right>$, suppressed by the reduced Planck mass $m_{\rm Pl} = 2.4 \times 10^{18}$ GeV.

\begin{figure}[htbp]
\centering
      \includegraphics[width=0.4\textwidth,angle=0]{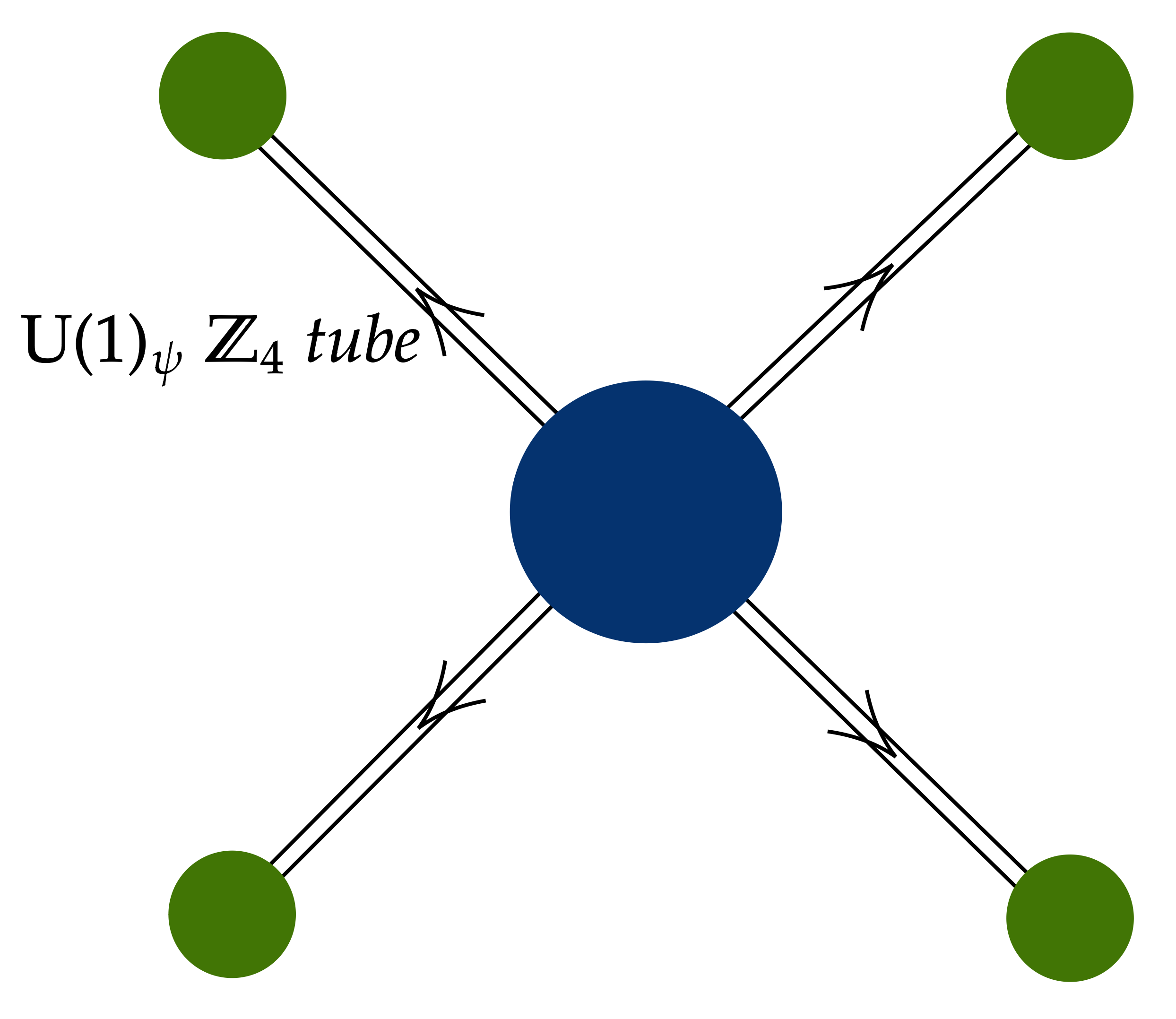}
       \caption{The flux of a monopole (in blue) corresponding to a $2\pi$ rotation along $U(1)_\psi$ is squeezed into four tubes. These tubes end on the minimal $U(1)_\psi$ antimonopoles, forming a ``fourfold" configuration.}

  \label{fig:quadcopter}
\end{figure}
We can realize an exotic finite energy configuration by combining a ``fourfold" monopole that carries one unit of $\psi$ magnetic flux and four elementary antimonopoles with each carrying $-1/4 \;\psi$ flux, as shown in Fig.~\ref{fig:quadcopter}. The stability of the fourfold monopole is contingent on the energy considerations and the potential barriers between different configurations. Depending on these factors, it is possible for the four antimonopoles to annihilate with the central monopole. Alternatively, the central monopole could split into four monopoles, which then connect to the four antimonopoles, thus forming four dumbbell-like structures. However, the rate of the central monopole splitting into four monopoles depends on the parameter space involved. For a related discussion involving the decay of higher charge monopoles, see Ref.~\cite{London:1985ve}.
\section{Domain wall bounded by necklace and strings from domain walls}
\label{sec:4}
\begin{figure}[htbp]
\centering
      \includegraphics[width=0.4\textwidth,angle=0]{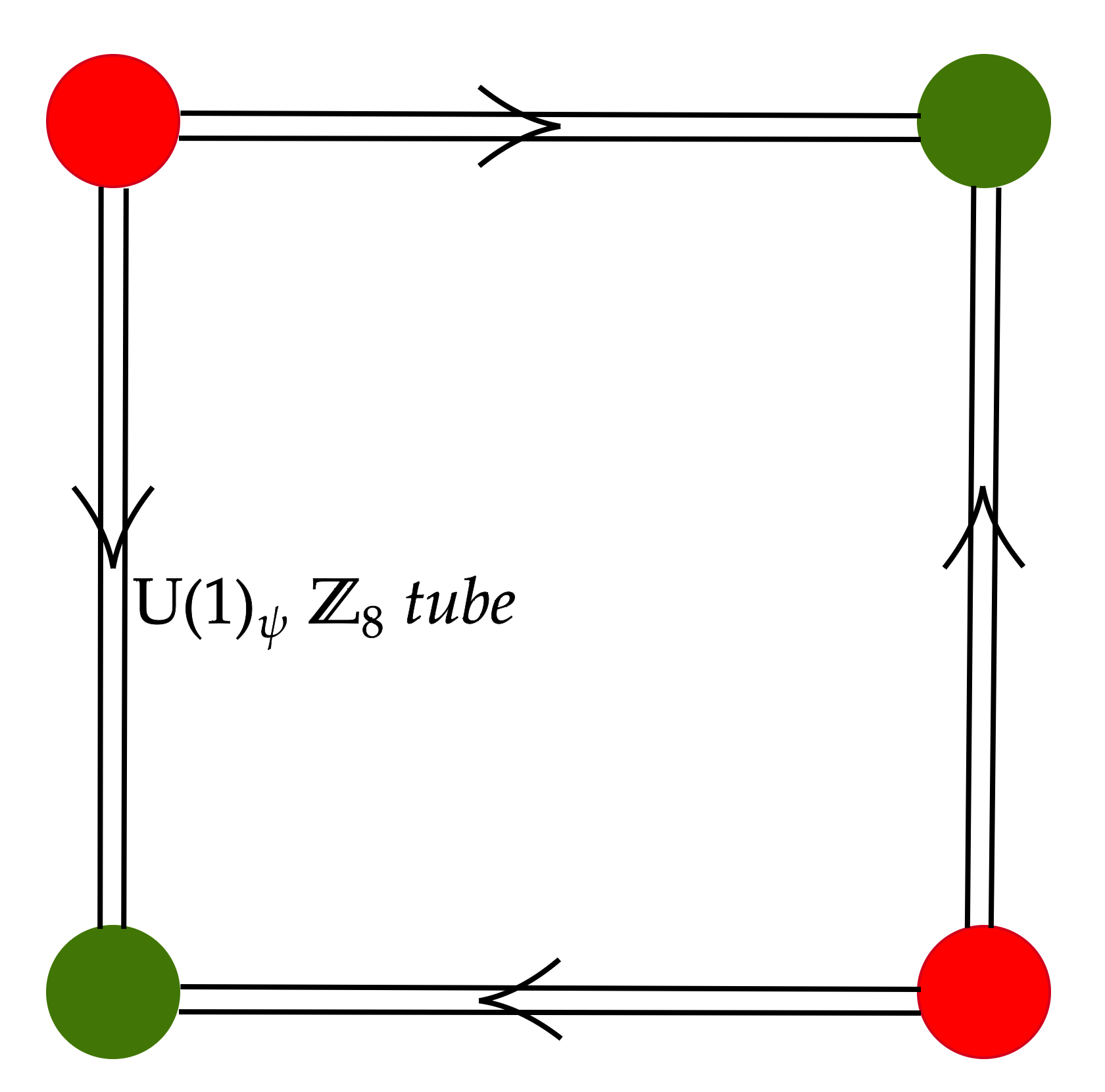}
       \caption{Necklace from the symmetry breaking $E_6\to SO(10)\times U(1)_\psi/\Z_4\to SO(10)\times \Z_8/\Z_4$. The symmetry breaking $U(1)_\psi\to \Z_8$ yields a flux tube carrying half the flux of the minimal $U(1)_\psi$ monopole.}
  \label{fig:necklacez8}
\end{figure}

The previous discussion provides a nice segue to a somewhat novel scenario for producing the metastable strings in which the $N_i$ matter fields acquire non-zero masses at renormalizable level from a non-zero vev of $351'$, which carries a $U(1)_\psi$ charge of $-8$. This breaks $U(1)_\psi$  to $\Z_8$, which produces a tube with flux $\psi/8$. With the $U(1)_\psi$ monopole carrying a flux
$\psi/4$, the confinement of these monopoles and antimonopoles now occurs via a necklace, as shown in Fig.~\ref{fig:necklacez8}.

\begin{figure}[h!]
\centering
      \includegraphics[width=0.4\textwidth,angle=0]{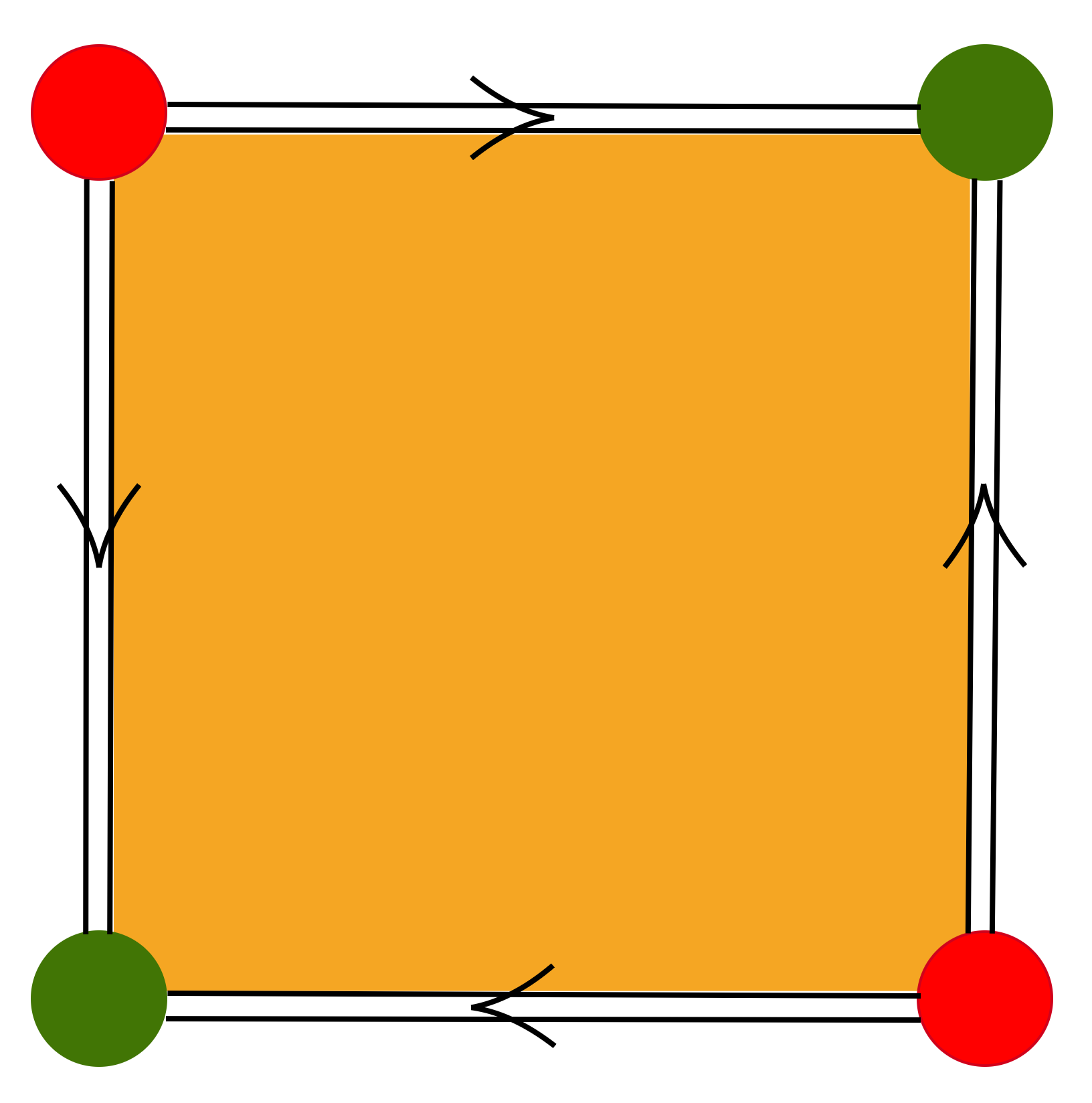}
       \caption{The breaking of $\Z_8\to \Z_4$ by the vev of $N$ results in a domain wall (in orange) bounded by a necklace configuration.}
  \label{fig:wallboundedbyneckalcez8}
\end{figure}

However, this is not the end of the story. A non-zero vev for $N$ is still required to provide masses to the fermions in the $SO(10)$ 10-plets, which breaks the $\Z_8$ symmetry to $\Z_4$, the center of $SO(10)$. Breaking of the $\Z_8$ symmetry to $\Z_4$ produces a domain wall, and so the end result appears to be the presence of a domain wall bounded by a necklace, as shown in Fig.~\ref{fig:wallboundedbyneckalcez8}.
Actually this is a reasonable conclusion since we have broken $E_6$ to $SO(10)$ as follows:
\begin{align}
\label{eq:e6-so10u1}
E_6 \to SO(10) \times U(1)_\psi \to SO(10)\times \Z_8 \to SO(10)           
\end{align}
Since $E_6$ is effectively broken to $SO(10)$, both the first and second homotopy groups of $E_6/SO(10)$ are trivial and no topologically stable structures are expected to be present.
And so the appearance of domain walls bounded by a necklace is compatible with this conclusion.

How then do we realize the metastable string scenario in this framework? If we assume that the primordial monopoles are inflated away, we are left with the following  symmetry breaking:
\begin{align}
\label{eq:break-u1-z8}
U(1)_\psi \to \Z_8 \to \Z_4
\end{align}
The first breaking of $U(1)_\psi$ in Eq.~(\ref{eq:break-u1-z8}) should produce $\Z_8$ strings, but the subsequent breaking of $\Z_8$ to $\Z_4$ creates domain walls. So we end up producing domain walls bounded by strings, and since $U(1)_\psi$ is effectively broken to $\Z_4$, there also should appear ``stable" $\Z_4$ strings. This happens from the collapse of the domain wall between the two $\Z_8$ strings. The ensuing $\Z_4$ string ends up being metastable due to the quantum mechanical pair production of $U(1)_\psi$ monopole-antimonopole pairs. Note that the collapse time $t_c$ for the $Z_8$ string-domain wall system that produces the $Z_4$ metastable string is of order $\mu/\sigma$ \cite{Vilenkin:1982ks,Vilenkin:1984ib,Martin:1996ea,Dunsky:2021tih}, where $\mu$ and $\sigma$ respectively denote the string mass per unit length and wall mass per unit area. Assuming a string scale of order $10^{16}$ GeV and the $N$ VEV lying somewhere between this value and $10$ TeV, the collapse time $t_c \lesssim 10^{-4}$ sec. The gravitational wave spectrum produced by the metastable strings is discussed in Section \ref{sec:gw}.

Finally, it is perhaps worth repeating that in the absence both of quantum mechanical tunneling and primordial $U(1)_\psi$ monopoles, the symmetry breaking $U(1)_\psi \longrightarrow Z_8 \longrightarrow Z_4$ (center of $SO(10)$) is expected to yield practically stable strings. We expect that the $Z_4$ strings will contribute in the usual way to the gravitational wave spectrum. 

This scenario may be contrasted with the symmetry breaking  $SU(2) \longrightarrow U(1)\longrightarrow Z_2 \longrightarrow 1$,  discussed in Ref.~\cite{Lazarides:2023iim}. A domain wall bounded by a necklace is also present in this case. However, since $SU(2)$ is a simply connected topological space, no strings remain at the end.


\section{Dark matter in $E_6$}
\label{sec:dm}
Dark matter (DM) candidates in $E_6$ models has been a subject of some discussion in recent years \cite{Hebbar:2017fit,Schwichtenberg:2017xhv,Barman:2017yzr,Bandyopadhyay:2019rja,Babu:2024ecl}. There is some overlap with the WIMP-like DM that arises in $SO(10)$ models supplemented by a $U(1)$ axion symmetry \cite{Holman:1982tb, Lazarides:2020frf, Okada:2022yvq,Lazarides:2022ezc}. We briefly summarize here the DM candidate associated with the breaking of $U(1)_\psi$ and the ensuing topological structure.
\begin{itemize}
    \item $U(1)_\psi$ breaking to $\Z_4$ through the vev of $N$. For $\langle N \rangle \sim M_{GUT}$ we find superheavy metastable strings that can explain the recent PTA measurements. A plausible DM candidate in this case would be an intermediate mass stable neutral particle present in the  vectorlike $SU(2)$ doublets \cite{Schwichtenberg:2017xhv, Lazarides:2020frf}. On the other hand, the slepton-like scalars ($\tilde{l}$ and $\tilde{\nu}$) from the scalar $SO(10)$ 16-plet are $\Z_2$-odd and therefore could be potential dark matter candidates. We can realize a two component dark matter scenario with one scalar and one fermion, if the masses of $N_i$ and the slepton-like scalars are tuned around the electroweak or TeV scale, whereas the masses of the vectorlike doublet fermions remain at the GUT scale \cite{Bandyopadhyay:2019rja}. The lightest neutral scalar or fermion is absolutely stable. The heavier component can decay into the lighter one only through a mixing with the charge neutral fermion fields from the vectorlike doublets, and this mixing is GUT scale suppressed. Thus, the heavier component is metastable and the lighter one is stable.
    \item For $\langle N \rangle \ll M_{GUT}$, the $U(1)_\psi$ string is essentially stable \cite{Witten:1984eb}, and depending on the vev of $N$, we could either have a doublet-singlet DM candidate \cite{Okada:2022yvq}, or the lightest $N_i$ as DM with mass in the keV range \cite{Hebbar:2017fit}.
    \item In the subsequent breaking of $SO(10)$ to the SM, we can obtain a metastable pseudo-Goldstone boson as dark matter particle when the singlet from $\overline{126}(-2)\in 351'$ acquires a vev ($\gtrsim 10^{11}$ GeV) to break the intermediate symmetry with rank five to the SM, and the neutrino-like scalar component of $16(1)\in 27$ has a vev around the electroweak scale ($10^2 - 10^5$ GeV). One of the would be Goldstone bosons is absorbed by the gauge boson associated with the broken diagonal generator. The other CP-odd scalar becomes a pseudo-Goldstone boson dark matter particle \cite{Abe:2021byq,Okada:2021qmi,Maji:2023fba}.
\end{itemize}
If $U(1)_\psi$ is broken in two steps to $\Z_4$, namely via $\Z_8$, we produce both walls bounded by necklaces as well as $\Z_4$ strings from the collapse of domain walls connecting two $\Z_8$ strings. Since the $351’$ vev is larger than the vev of $N$, we expect the fermions $N_i$ to be heavier than the $10$-plet fermions in $SO(10)$. In this case the neutral fermions in the latter multiplets provide dark matter in the $10^{12}$ GeV mass range.
\section{Gravitational waves from metastable and stable strings}
\label{sec:gw}
The dumbells, formed during the decay of the metastable strings carrying $U(1)_\psi$ flux, carry no unconfined flux. The subsequent breaking of the diagonal generator orthogonal to the SM hypercharge can be done with the vev of a tensorial 126-plet or a spinorial 16-plet scalar. The former produces topologically stable $\Z_2$ strings, whereas the latter can produce metastable strings depending on the symmetry breaking pattern. The gravitational waves from metastable strings have been extensively studied in light of the pulsar timing array experiments \cite{NANOGrav:2023hvm, Buchmuller:2023aus,Antusch:2023zjk,Lazarides:2023rqf,Maji:2023fhv,Ahmed:2023rky,Afzal:2023cyp,Afzal:2023kqs,Ahmed:2023pjl,Lazarides:2023bjd,Pallis:2024mip}. In this section we estimate the gravitational wave spectra if both the stable and metastable strings are present. 

The stochastic gravitational wave background is given by \cite{Leblond:2009fq, Olmez:2010bi}
 \begin{align}\label{eq:GWs}
\Omega_{\rm GW}(f) = \frac{4\pi^2}{3H_0^2}f^3\int_{z_{\rm min}}^{z_F} dz \int dl \, h^2(f,l,z)\frac{d^2R}{dz \, dl} \ ,
\end{align}
where $h(f,l,z)$ is the waveform of the burst, and the burst rate per unit space-time volume is expressed as
\begin{align}
\frac{d^2R}{dz \, dl} = N_b H_0^{-3}\phi_V(z) \frac{n(l,t)}{T(1+z)}\Delta(f,l,z) .
\end{align}
Here, $H_0$ is the present day Hubble parameter, $N_b$ is the average number of burst events in a oscillation period $T$, $n(l,t)$ the number distribution, $\Delta(f,l,z)$ is the observable fraction of the bursts \cite{Damour:2001bk,Leblond:2009fq, Olmez:2010bi}, and 
\begin{align}
\phi_V(z) = \frac{4\pi H_0^3 r^2(z)}{(1+z)^3{H}(z)},
\end{align}
with $H(z)$ being the Hubble parameter and $r(z)$ the proper distance at the redshift $z$. In Eq.~\eqref{eq:GWs}, $z_F$ denotes the redshift at the network formation, and the lower limit $z_{min}$ is chosen to remove the recent infrequent bursts from the background \cite{Damour:2001bk}, or it corresponds the redshift when the metastable network disappears.

\begin{figure}[ht]
\centering
\includegraphics[width=0.8\textwidth]{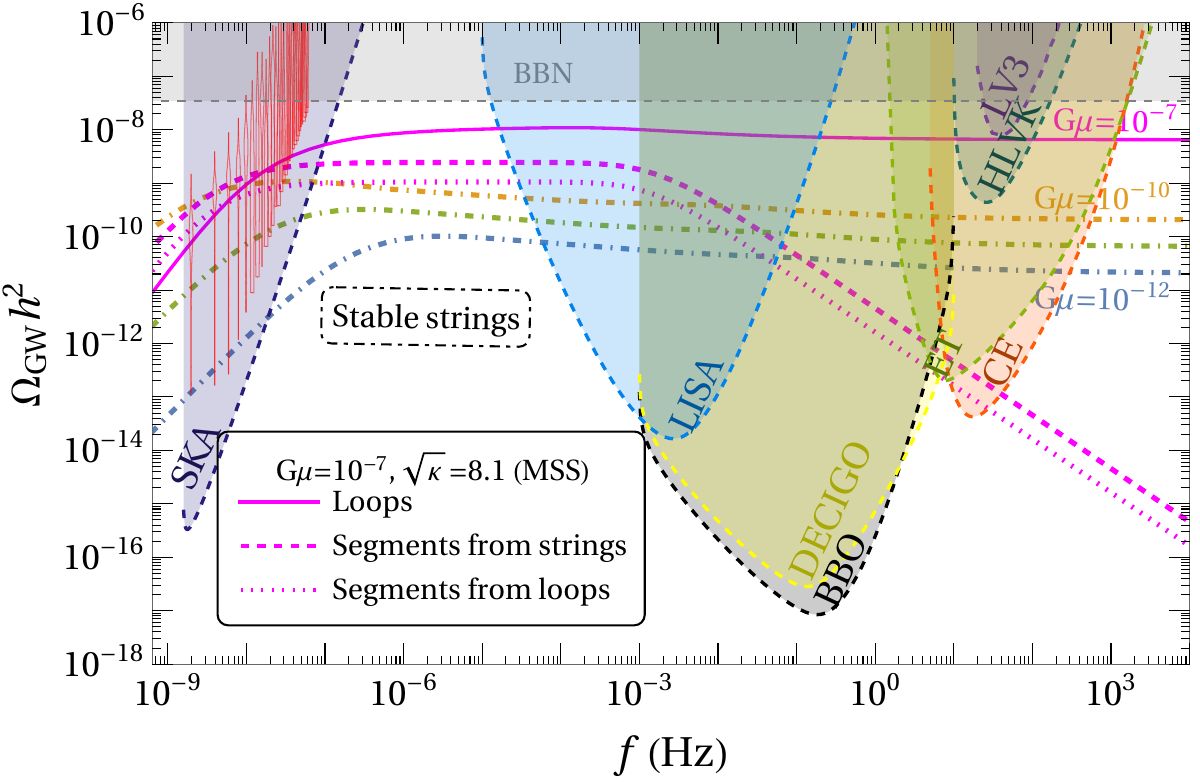}
\caption{Gravitational wave background from the metastable cosmic strings with $G\mu=10^{-7}$, $\sqrt{\kappa} = 8.1$, and separately from intermediate scale stable cosmic strings with $G\mu=10^{-12}, 10^{-11}, 10^{-10}$ (dot-dashed curves). The former has contributions from the loops (solid magenta curve) as well as segments from the decay of the long strings (dashed magenta) and loops (dotted magenta).  Clearly, the contribution from the stable strings is subdominant. However, their contribution can be important for the lowest one or two frequency bins for $10^{-10}\gtrsim G\mu\gtrsim 10^{-11}$. The violin plots depict the posteriors of Hellings-Downs correlated \cite{Hellings:1983fr} free spectra for NANOGrav data \cite{NANOGrav:2023hvm}. The constraints from Big Bang Nucleosynthesis (BBN) \cite{Mangano:2011ar} and LIGO-VIRGO data \cite{LIGOScientific:2021nrg} are also shown. Several proposed experiments, namely, HLVK \cite{KAGRA:2013rdx}, CE \cite{Regimbau:2016ike}, ET \cite{Mentasti:2020yyd}, DECIGO \cite{Sato_2017}, BBO \cite{Crowder:2005nr, Corbin:2005ny}, LISA \cite{Bartolo:2016ami, amaroseoane2017laser} and SKA \cite{5136190, Janssen:2014dka} can probe the gravitational wave spectra as dictated by their respective power-law integrated sensitivity curves \cite{Thrane:2013oya, Schmitz:2020syl}.}
\label{fig:GWs}
\end{figure}

 We assume that these two types of strings enter the scaling regime \cite{Vachaspati:1984gt,Vilenkin:2000jqa,Vanchurin:2005pa,Ringeval:2005kr,Olum:2006ix,Blanco-Pillado:2013qja,Blanco-Pillado:2017oxo,Cui:2018rwi} independently with loop distributions given in the Refs.~ \cite{Blanco-Pillado:2013qja,Blanco-Pillado:2017oxo,LIGOScientific:2021nrg,Buchmuller:2021mbb}. The superheavy metastable strings confine flux associated with $U(1)_\psi$ breaking, whereas the intermediate scale stable strings carry flux associated with the broken generator $U(1)_\chi$ in $SO(10)$. The former will see the latter as a vacuum with higher energy, and therefore, we neglect their interactions. Indeed, this is an important topic that requires detailed numerical simulations. We estimate the gravitational wave background as a consequence of the unresolvable bursts, dominantly from the cusps \cite{Damour:2001bk, Olmez:2010bi, Auclair:2019wcv, Cui:2019kkd} for the loops. For more details see Refs.~\cite{Lazarides:2022ezc, Lazarides:2023rqf} and references therein. The monopole-antimonopole connected by the string segments (Fig.~\ref{fig:dumbellonefoure6monopole}) do not carry any unconfined flux. The subhorizon segments from the decay of long strings as well as the loops oscillate and radiate gravitational waves. We estimate the gravitational waves assuming the straight string approximation and follow the prescription of Ref.~\cite{Leblond:2009fq}. We use the number distribution of segments from the loops from Ref.~\cite{Buchmuller:2021mbb}. The gravitational power radiated from a segment is $P_{\rm seg}=\tilde{\Gamma} G\mu^2$ and we assume $\tilde{\Gamma} = 50$ as in \cite{NANOGrav:2023hvm}.
 
Fig.~\ref{fig:GWs} shows the gravitational wave backgrounds from the metastable cosmic strings with $G\mu=10^{-7}$ and the stable intermediate scale cosmic strings with $G\mu=10^{-12}-10^{-10}$. The metastable strings with a metastability factor $\sqrt{\kappa}\approx 8$ ($\kappa=m_M^2/\mu$ with $m_M$ being the monopole mass) and $G\mu=10^{-7}$, lie in the $68\%$ Bayesian credible interval \cite{NANOGrav:2023hvm}. A higher $G\mu$ value with a suitable $\sqrt{\kappa}$ could provide a better fit to the NANOGrav data, though it would contradict the bound from LIGO-VIRGO third run (LV3) data \cite{LIGOScientific:2021nrg} around the decaHertz frequencies. In this case, the strings need to experience some $e$-foldings of inflation, or we need to invoke a non-standard cosmology with early matter domination. It is worth mentioning that for $G\mu > 10^{-7}$ \cite{Lazarides:2023ksx,Lazarides:2023rqf}, if the metastable strings experience some $e$-foldings of primordial inflation and therefore enter the scaling regime at a late time $t_F$ after their horizon reentry, the gravitational wave spectrum decreases as a power-law $f^{-1/3}$ at high frequencies and satisfies the constraint set by the LIGO-VIRGO data \cite{LIGOScientific:2021nrg}. The gravitational waves from the stable strings remain subdominant, of course. However, for $10^{-10}\gtrsim G\mu\gtrsim 10^{-11}$ their contribution can be important for the lowest one or two frequency bins of PTA data.
Several proposed experiments, namely, HLVK \cite{KAGRA:2013rdx}, CE \cite{Regimbau:2016ike}, ET \cite{Mentasti:2020yyd}, DECIGO \cite{Sato_2017}, BBO \cite{Crowder:2005nr, Corbin:2005ny}, LISA \cite{Bartolo:2016ami, amaroseoane2017laser} and SKA \cite{5136190, Janssen:2014dka} can probe the gravitational wave spectra as dictated by their respective power-law integrated sensitivity curves \cite{Thrane:2013oya, Schmitz:2020syl}.

Before closing, let us note that the metastable strings discussed here can carry both electrically charged and electrically neutral fermion zero modes. The stable intermediate scale string from $SO(10)$ breaking carries right handed neutrino zero modes because the SM singlet vev in the tensorial 126-dimensional scalar couples only to the SM-singlet right-handed neutrino. The impact of these neutral zero modes on the gravitational wave spectrum has recently been discussed in Ref.~\cite{Afzal:2023kqs}.
\section{Conclusion}
We have explored a variety of composite topological structures that appear in the spontaneous breaking of $E_6$ grand unification symmetry to the Standard Model. We focused exclusively on structures arising from the symmetry breaking of $E_6$ via its maximal subgroup $SO(10) \times U(1)_\psi$, and the composite topological structures we identify provide a test of $E_6$ grand unification if they appeared in the early universe. We identify a novel scenario for realizing metastable strings in the $E_6$ model. The breaking of $U(1)_\psi$ symmetry to $\Z_4$ (center of $SO(10)$) in this case proceeds in two steps and yields the metastable strings from the collapse of domain walls.

A remnant $\Z_2$ gauge symmetry from the $SO(10)$ sector in $E_6$ is essential in ensuring the stability of a viable dark matter particle. We also identify a dark matter scenario consisting of keV mass sterile neutrino. The gravitational spectrum generated by superheavy metastable cosmic strings and intermediate scale topologically stable strings is also described.

Finally, a few remarks about topological structures and primordial inflation may be in order here. For instance, in order to implement the metastable string scenario the primordial monopole is inflated away but certainly not the string, even though their symmetry breaking scales are close to each other. This can be realized in models of hybrid inflation by suitably adjusting the parameters associated  with the appropriate `waterfall' fields \cite{Lazarides:2023rqf,Moursy:2024hll}. The $E_6$ model discussed here offers a variety of symmetry breaking scenarios and associated topological structures that should be testable in the ongoing and future experiments.
\section*{Acknowledgments}
R.M. is supported by Institute for Basic Science under the project code: IBS-R018-D3. A.T. is partially supported by the Bartol Research Institute, University of Delaware.
\appendix
%
%
%
%
%
%
%
%
%
%
%
%
%

\bibliographystyle{JHEP}
\bibliography{topological.bib}

\providecommand{\href}[2]{#2}\begingroup\raggedright\begin{thebibliography}{10}

\bibitem{Georgi:1974my}
H.~Georgi, \emph{{The State of the Art\textemdash{}Gauge Theories}}, \href{https://doi.org/10.1063/1.2947450}{\emph{AIP Conf. Proc.} {\bfseries 23} (1975) 575}.

\bibitem{Fritzsch:1974nn}
H.~Fritzsch and P.~Minkowski, \emph{{Unified Interactions of Leptons and Hadrons}}, \href{https://doi.org/10.1016/0003-4916(75)90211-0}{\emph{Annals Phys.} {\bfseries 93} (1975) 193}.

\bibitem{Kibble:1982ae}
T.W.B.~Kibble, G.~Lazarides and Q.~Shafi, \emph{{Strings in SO(10)}}, \href{https://doi.org/10.1016/0370-2693(82)90829-2}{\emph{Phys. Lett. B} {\bfseries 113} (1982) 237}.

\bibitem{Lazarides:1982tw}
G.~Lazarides and Q.~Shafi, \emph{{Axion Models with No Domain Wall Problem}}, \href{https://doi.org/10.1016/0370-2693(82)90506-8}{\emph{Phys. Lett. B} {\bfseries 115} (1982) 21}.

\bibitem{Holman:1982tb}
R.~Holman, G.~Lazarides and Q.~Shafi, \emph{{Axions and the Dark Matter of the Universe}}, \href{https://doi.org/10.1103/PhysRevD.27.995}{\emph{Phys. Rev. D} {\bfseries 27} (1983) 995}.

\bibitem{Lazarides:2020frf}
G.~Lazarides and Q.~Shafi, \emph{{Axion Model with Intermediate Scale Fermionic Dark Matter}}, \href{https://doi.org/10.1016/j.physletb.2020.135603}{\emph{Phys. Lett. B} {\bfseries 807} (2020) 135603} [\href{https://arxiv.org/abs/2004.11560}{{\ttfamily 2004.11560}}].

\bibitem{Kadastik:2009dj}
M.~Kadastik, K.~Kannike and M.~Raidal, \emph{{Matter parity as the origin of scalar Dark Matter}}, \href{https://doi.org/10.1103/PhysRevD.81.015002}{\emph{Phys. Rev. D} {\bfseries 81} (2010) 015002} [\href{https://arxiv.org/abs/0903.2475}{{\ttfamily 0903.2475}}].

\bibitem{Mambrini:2015vna}
Y.~Mambrini, N.~Nagata, K.A.~Olive, J.~Quevillon and J.~Zheng, \emph{{Dark matter and gauge coupling unification in nonsupersymmetric SO(10) grand unified models}}, \href{https://doi.org/10.1103/PhysRevD.91.095010}{\emph{Phys. Rev. D} {\bfseries 91} (2015) 095010} [\href{https://arxiv.org/abs/1502.06929}{{\ttfamily 1502.06929}}].

\bibitem{Boucenna:2015sdg}
S.M.~Boucenna, M.B.~Krauss and E.~Nardi, \emph{{Dark matter from the vector of SO (10)}}, \href{https://doi.org/10.1016/j.physletb.2016.02.008}{\emph{Phys. Lett. B} {\bfseries 755} (2016) 168} [\href{https://arxiv.org/abs/1511.02524}{{\ttfamily 1511.02524}}].

\bibitem{Ferrari:2018rey}
S.~Ferrari, T.~Hambye, J.~Heeck and M.H.G.~Tytgat, \emph{{SO(10) paths to dark matter}}, \href{https://doi.org/10.1103/PhysRevD.99.055032}{\emph{Phys. Rev. D} {\bfseries 99} (2019) 055032} [\href{https://arxiv.org/abs/1811.07910}{{\ttfamily 1811.07910}}].

\bibitem{Bhattacharyya:2021lgr}
S.~Bhattacharyya and A.~Datta, \emph{{Phenomenology of an E6 inspired extension of the Standard Model: Higgs sector}}, \href{https://doi.org/10.1103/PhysRevD.105.075021}{\emph{Phys. Rev. D} {\bfseries 105} (2022) 075021} [\href{https://arxiv.org/abs/2109.08524}{{\ttfamily 2109.08524}}].

\bibitem{Lazarides:2022spe}
G.~Lazarides, R.~Maji, R.~Roshan and Q.~Shafi, \emph{{Heavier W boson, dark matter, and gravitational waves from strings in an SO(10) axion model}}, \href{https://doi.org/10.1103/PhysRevD.106.055009}{\emph{Phys. Rev. D} {\bfseries 106} (2022) 055009} [\href{https://arxiv.org/abs/2205.04824}{{\ttfamily 2205.04824}}].

\bibitem{Lazarides:2022ezc}
G.~Lazarides, R.~Maji, R.~Roshan and Q.~Shafi, \emph{{A predictive SO(10) model}}, \href{https://doi.org/10.1088/1475-7516/2022/12/009}{\emph{JCAP} {\bfseries 12} (2022) 009} [\href{https://arxiv.org/abs/2210.03710}{{\ttfamily 2210.03710}}].

\bibitem{Sahu:2022rwq}
P.~Sahu, A.~Bhatta, R.~Mohanta, S.~Singirala and S.~Patra, \emph{{Flavour anomalies and dark matter assisted unification in SO(10) GUT}}, \href{https://doi.org/10.1007/JHEP11(2022)029}{\emph{JHEP} {\bfseries 11} (2022) 029} [\href{https://arxiv.org/abs/2204.06392}{{\ttfamily 2204.06392}}].

\bibitem{Bhattacharyya:2022trp}
S.~Bhattacharyya and A.~Datta, \emph{{Dark matter perspective of left-right symmetric gauge model}}, \href{https://doi.org/10.1016/j.nuclphysb.2023.116197}{\emph{Nucl. Phys. B} {\bfseries 991} (2023) 116197} [\href{https://arxiv.org/abs/2206.13105}{{\ttfamily 2206.13105}}].

\bibitem{Okada:2022yvq}
N.~Okada, D.~Raut and Q.~Shafi, \emph{{Axions, WIMPs, proton decay and observable r in SO(10)}}, \href{https://doi.org/10.1140/epjc/s10052-023-11378-1}{\emph{Eur. Phys. J. C} {\bfseries 83} (2023) 273} [\href{https://arxiv.org/abs/2207.10538}{{\ttfamily 2207.10538}}].

\bibitem{Kibble:1982dd}
T.W.B.~Kibble, G.~Lazarides and Q.~Shafi, \emph{{Walls Bounded by Strings}}, \href{https://doi.org/10.1103/PhysRevD.26.435}{\emph{Phys. Rev. D} {\bfseries 26} (1982) 435}.

\bibitem{Lazarides:1985my}
G.~Lazarides and Q.~Shafi, \emph{{Superconducting Membranes}}, \href{https://doi.org/10.1016/0370-2693(85)90246-1}{\emph{Phys. Lett. B} {\bfseries 159} (1985) 261}.

\bibitem{Chang:1983fu}
D.~Chang, R.N.~Mohapatra and M.K.~Parida, \emph{{Decoupling Parity and $SU(2)_R$ Breaking Scales: A New Approach to Left-Right Symmetric Models}}, \href{https://doi.org/10.1103/PhysRevLett.52.1072}{\emph{Phys. Rev. Lett.} {\bfseries 52} (1984) 1072}.

\bibitem{Makinen:2018ltj}
J.T.~M\"akinen, V.V.~Dmitriev, J.~Nissinen, J.~Rysti, G.E.~Volovik, A.N.~Yudin et~al., \emph{{Half-quantum vortices and walls bounded by strings in the polar-distorted phases of topological superfluid$^{3}$He}}, \href{https://doi.org/10.1038/s41467-018-08204-8}{\emph{Nature Commun.} {\bfseries 10} (2019) 237} [\href{https://arxiv.org/abs/1807.04328}{{\ttfamily 1807.04328}}].

\bibitem{Dunsky:2021tih}
D.I.~Dunsky, A.~Ghoshal, H.~Murayama, Y.~Sakakihara and G.~White, \emph{{GUTs, hybrid topological defects, and gravitational waves}}, \href{https://doi.org/10.1103/PhysRevD.106.075030}{\emph{Phys. Rev. D} {\bfseries 106} (2022) 075030} [\href{https://arxiv.org/abs/2111.08750}{{\ttfamily 2111.08750}}].

\bibitem{Maji:2023fba}
R.~Maji, W.-I.~Park and Q.~Shafi, \emph{{Gravitational waves from walls bounded by strings in SO(10) model of pseudo-Goldstone dark matter}}, \href{https://doi.org/10.1016/j.physletb.2023.138127}{\emph{Phys. Lett. B} {\bfseries 845} (2023) 138127} [\href{https://arxiv.org/abs/2305.11775}{{\ttfamily 2305.11775}}].

\bibitem{Eto:2023aqr}
M.~Eto, T.~Hiramatsu, I.~Saito and Y.~Sakakihara, \emph{{String-wall composites winding around a torus knot vacuum in an axionlike model}}, \href{https://doi.org/10.1103/PhysRevD.108.116004}{\emph{Phys. Rev. D} {\bfseries 108} (2023) 116004} [\href{https://arxiv.org/abs/2309.04248}{{\ttfamily 2309.04248}}].

\bibitem{Roshan:2024qnv}
R.~Roshan and G.~White, \emph{{Using gravitational waves to see the first second of the Universe}},  \href{https://arxiv.org/abs/2401.04388}{{\ttfamily 2401.04388}}.

\bibitem{Fu:2024rsm}
B.~Fu, A.~Ghoshal, S.F.~King and M.H.~Rahat, \emph{{Type-I two-Higgs-doublet model and gravitational waves from domain walls bounded by strings}},  \href{https://arxiv.org/abs/2404.16931}{{\ttfamily 2404.16931}}.

\bibitem{Hamada:2024dan}
Y.~Hamada and W.~Nakano, \emph{{Gravitational wave spectrum from expanding string loops on domain walls: Implication to nano-hertz pulsar timing array signal}},  \href{https://arxiv.org/abs/2405.09599}{{\ttfamily 2405.09599}}.

\bibitem{Lazarides:2023iim}
G.~Lazarides, Q.~Shafi and A.~Tiwari, \emph{{Composite topological structures in SO(10)}}, \href{https://doi.org/10.1007/JHEP05(2023)119}{\emph{JHEP} {\bfseries 05} (2023) 119} [\href{https://arxiv.org/abs/2303.15159}{{\ttfamily 2303.15159}}].

\bibitem{Hindmarsh:1985xc}
M.~Hindmarsh and T.W.B.~Kibble, \emph{{Beads on Strings}}, \href{https://doi.org/10.1103/PhysRevLett.55.2398}{\emph{Phys. Rev. Lett.} {\bfseries 55} (1985) 2398}.

\bibitem{Aryal:1987sn}
M.~Aryal and A.E.~Everett, \emph{{Properties of $Z_2$ Strings}}, \href{https://doi.org/10.1103/PhysRevD.35.3105}{\emph{Phys. Rev. D} {\bfseries 35} (1987) 3105}.

\bibitem{Kibble:2015twa}
T.W.B.~Kibble and T.~Vachaspati, \emph{{Monopoles on strings}}, \href{https://doi.org/10.1088/0954-3899/42/9/094002}{\emph{J. Phys. G} {\bfseries 42} (2015) 094002} [\href{https://arxiv.org/abs/1506.02022}{{\ttfamily 1506.02022}}].

\bibitem{Gursey:1975ki}
F.~Gursey, P.~Ramond and P.~Sikivie, \emph{{A Universal Gauge Theory Model Based on E6}}, \href{https://doi.org/10.1016/0370-2693(76)90417-2}{\emph{Phys. Lett. B} {\bfseries 60} (1976) 177}.

\bibitem{Achiman:1978vg}
Y.~Achiman and B.~Stech, \emph{{Quark Lepton Symmetry and Mass Scales in an E6 Unified Gauge Model}}, \href{https://doi.org/10.1016/0370-2693(78)90584-1}{\emph{Phys. Lett. B} {\bfseries 77} (1978) 389}.

\bibitem{Shafi:1978gg}
Q.~Shafi, \emph{{E(6) as a Unifying Gauge Symmetry}}, \href{https://doi.org/10.1016/0370-2693(78)90248-4}{\emph{Phys. Lett. B} {\bfseries 79} (1978) 301}.

\bibitem{Lazarides:2019xai}
G.~Lazarides and Q.~Shafi, \emph{{Monopoles, Strings, and Necklaces in $SO(10)$ and $E_6$}}, \href{https://doi.org/10.1007/JHEP10(2019)193}{\emph{JHEP} {\bfseries 10} (2019) 193} [\href{https://arxiv.org/abs/1904.06880}{{\ttfamily 1904.06880}}].

\bibitem{Witten:1984eb}
E.~Witten, \emph{{Superconducting Strings}}, \href{https://doi.org/10.1016/0550-3213(85)90022-7}{\emph{Nucl. Phys. B} {\bfseries 249} (1985) 557}.

\bibitem{NANOGrav:2023hvm}
{\scshape NANOGrav} collaboration, \emph{{The NANOGrav 15 yr Data Set: Search for Signals from New Physics}}, \href{https://doi.org/10.3847/2041-8213/acdc91}{\emph{Astrophys. J. Lett.} {\bfseries 951} (2023) L11} [\href{https://arxiv.org/abs/2306.16219}{{\ttfamily 2306.16219}}].

\bibitem{Buchmuller:2023aus}
W.~Buchmuller, V.~Domcke and K.~Schmitz, \emph{{Metastable cosmic strings}}, \href{https://doi.org/10.1088/1475-7516/2023/11/020}{\emph{JCAP} {\bfseries 11} (2023) 020} [\href{https://arxiv.org/abs/2307.04691}{{\ttfamily 2307.04691}}].

\bibitem{Antusch:2023zjk}
S.~Antusch, K.~Hinze, S.~Saad and J.~Steiner, \emph{{Singling out SO(10) GUT models using recent PTA results}}, \href{https://doi.org/10.1103/PhysRevD.108.095053}{\emph{Phys. Rev. D} {\bfseries 108} (2023) 095053} [\href{https://arxiv.org/abs/2307.04595}{{\ttfamily 2307.04595}}].

\bibitem{Lazarides:2023rqf}
G.~Lazarides, R.~Maji, A.~Moursy and Q.~Shafi, \emph{{Inflation, superheavy metastable strings and gravitational waves in non-supersymmetric flipped SU(5)}}, \href{https://doi.org/10.1088/1475-7516/2024/03/006}{\emph{JCAP} {\bfseries 03} (2024) 006} [\href{https://arxiv.org/abs/2308.07094}{{\ttfamily 2308.07094}}].

\bibitem{Maji:2023fhv}
R.~Maji and W.-I.~Park, \emph{{Supersymmetric U(1)B-L flat direction and NANOGrav 15 year data}}, \href{https://doi.org/10.1088/1475-7516/2024/01/015}{\emph{JCAP} {\bfseries 01} (2024) 015} [\href{https://arxiv.org/abs/2308.11439}{{\ttfamily 2308.11439}}].

\bibitem{Ahmed:2023rky}
W.~Ahmed, M.U.~Rehman and U.~Zubair, \emph{{Probing stochastic gravitational wave background from SU(5)~\texttimes{}~U(1)$_{\chi}$ strings in light of NANOGrav 15-year data}}, \href{https://doi.org/10.1088/1475-7516/2024/01/049}{\emph{JCAP} {\bfseries 01} (2024) 049} [\href{https://arxiv.org/abs/2308.09125}{{\ttfamily 2308.09125}}].

\bibitem{Afzal:2023cyp}
A.~Afzal, M.~Mehmood, M.U.~Rehman and Q.~Shafi, \emph{{Supersymmetric hybrid inflation and metastable cosmic strings in $SU(4)_c \times SU(2)_L \times U(1)_R$}},  \href{https://arxiv.org/abs/2308.11410}{{\ttfamily 2308.11410}}.

\bibitem{Afzal:2023kqs}
A.~Afzal, Q.~Shafi and A.~Tiwari, \emph{{Gravitational wave emission from metastable current-carrying strings in E6}}, \href{https://doi.org/10.1016/j.physletb.2024.138516}{\emph{Phys. Lett. B} {\bfseries 850} (2024) 138516} [\href{https://arxiv.org/abs/2311.05564}{{\ttfamily 2311.05564}}].

\bibitem{Ahmed:2023pjl}
W.~Ahmed, T.A.~Chowdhury, S.~Nasri and S.~Saad, \emph{{Gravitational waves from metastable cosmic strings in the Pati-Salam model in light of new pulsar timing array data}}, \href{https://doi.org/10.1103/PhysRevD.109.015008}{\emph{Phys. Rev. D} {\bfseries 109} (2024) 015008} [\href{https://arxiv.org/abs/2308.13248}{{\ttfamily 2308.13248}}].

\bibitem{Lazarides:2023bjd}
G.~Lazarides and C.~Pallis, \emph{{Probing the supersymmetry-mass scale with F-term hybrid inflation}}, \href{https://doi.org/10.1103/PhysRevD.108.095055}{\emph{Phys. Rev. D} {\bfseries 108} (2023) 095055} [\href{https://arxiv.org/abs/2309.04848}{{\ttfamily 2309.04848}}].

\bibitem{Pallis:2024mip}
C.~Pallis, \emph{{PeV-Scale SUSY and Cosmic Strings from F-term Hybrid Inflation}}, \href{https://doi.org/10.3390/universe10050211}{\emph{Universe} {\bfseries 10} (2024) 5} [\href{https://arxiv.org/abs/2403.09385}{{\ttfamily 2403.09385}}].

\bibitem{NANOGrav:2023gor}
{\scshape NANOGrav} collaboration, \emph{{The NANOGrav 15 yr Data Set: Evidence for a Gravitational-wave Background}}, \href{https://doi.org/10.3847/2041-8213/acdac6}{\emph{Astrophys. J. Lett.} {\bfseries 951} (2023) L8} [\href{https://arxiv.org/abs/2306.16213}{{\ttfamily 2306.16213}}].

\bibitem{EPTA:2023fyk}
{\scshape EPTA, InPTA:} collaboration, \emph{{The second data release from the European Pulsar Timing Array - III. Search for gravitational wave signals}}, \href{https://doi.org/10.1051/0004-6361/202346844}{\emph{Astron. Astrophys.} {\bfseries 678} (2023) A50} [\href{https://arxiv.org/abs/2306.16214}{{\ttfamily 2306.16214}}].

\bibitem{Reardon:2023gzh}
D.J.~Reardon et~al., \emph{{Search for an Isotropic Gravitational-wave Background with the Parkes Pulsar Timing Array}}, \href{https://doi.org/10.3847/2041-8213/acdd02}{\emph{Astrophys. J. Lett.} {\bfseries 951} (2023) L6} [\href{https://arxiv.org/abs/2306.16215}{{\ttfamily 2306.16215}}].

\bibitem{Xu:2023wog}
H.~Xu et~al., \emph{{Searching for the Nano-Hertz Stochastic Gravitational Wave Background with the Chinese Pulsar Timing Array Data Release I}}, \href{https://doi.org/10.1088/1674-4527/acdfa5}{\emph{Res. Astron. Astrophys.} {\bfseries 23} (2023) 075024} [\href{https://arxiv.org/abs/2306.16216}{{\ttfamily 2306.16216}}].

\bibitem{Moursy:2024hll}
A.~Moursy and Q.~Shafi, \emph{{Primordial monopoles, black holes and gravitational waves}},  \href{https://arxiv.org/abs/2405.04397}{{\ttfamily 2405.04397}}.

\bibitem{London:1985ve}
D.~London, \emph{{Is the Doubly Charged Monopole Stable?}}, \href{https://doi.org/10.1103/PhysRevD.33.3075}{\emph{Phys. Rev. D} {\bfseries 33} (1986) 3075}.

\bibitem{Vilenkin:1982ks}
A.~Vilenkin and A.E.~Everett, \emph{{Cosmic Strings and Domain Walls in Models with Goldstone and PseudoGoldstone Bosons}}, \href{https://doi.org/10.1103/PhysRevLett.48.1867}{\emph{Phys. Rev. Lett.} {\bfseries 48} (1982) 1867}.

\bibitem{Vilenkin:1984ib}
A.~Vilenkin, \emph{{Cosmic Strings and Domain Walls}}, \href{https://doi.org/10.1016/0370-1573(85)90033-X}{\emph{Phys. Rept.} {\bfseries 121} (1985) 263}.

\bibitem{Martin:1996ea}
X.~Martin and A.~Vilenkin, \emph{{Gravitational wave background from hybrid topological defects}}, \href{https://doi.org/10.1103/PhysRevLett.77.2879}{\emph{Phys. Rev. Lett.} {\bfseries 77} (1996) 2879} [\href{https://arxiv.org/abs/astro-ph/9606022}{{\ttfamily astro-ph/9606022}}].

\bibitem{Hebbar:2017fit}
A.~Hebbar, G.~Lazarides and Q.~Shafi, \emph{{Light sterile neutrinos, dark matter, and new resonances in a U(1) extension of the MSSM}}, \href{https://doi.org/10.1103/PhysRevD.96.055026}{\emph{Phys. Rev. D} {\bfseries 96} (2017) 055026} [\href{https://arxiv.org/abs/1706.09630}{{\ttfamily 1706.09630}}].

\bibitem{Schwichtenberg:2017xhv}
J.~Schwichtenberg, \emph{{Dark matter in E$_{6}$ Grand unification}}, \href{https://doi.org/10.1007/JHEP02(2018)016}{\emph{JHEP} {\bfseries 02} (2018) 016} [\href{https://arxiv.org/abs/1704.04219}{{\ttfamily 1704.04219}}].

\bibitem{Barman:2017yzr}
B.~Barman, S.~Bhattacharya, S.K.~Patra and J.~Chakrabortty, \emph{{Non-Abelian Vector Boson Dark Matter, its Unified Route and signatures at the LHC}}, \href{https://doi.org/10.1088/1475-7516/2017/12/021}{\emph{JCAP} {\bfseries 12} (2017) 021} [\href{https://arxiv.org/abs/1704.04945}{{\ttfamily 1704.04945}}].

\bibitem{Bandyopadhyay:2019rja}
T.~Bandyopadhyay and R.~Maji, \emph{{The E6 route to multicomponent dark matter}},  \href{https://arxiv.org/abs/1911.13298}{{\ttfamily 1911.13298}}.

\bibitem{Babu:2024ecl}
K.S.~Babu, B.~Bajc and V.~Susi\v{c}, \emph{{A realistic theory of $\mathrm{E_{6}}$ unification through novel intermediate symmetries}},  \href{https://arxiv.org/abs/2403.20278}{{\ttfamily 2403.20278}}.

\bibitem{Abe:2021byq}
Y.~Abe, T.~Toma, K.~Tsumura and N.~Yamatsu, \emph{{Pseudo-Nambu-Goldstone dark matter model inspired by grand unification}}, \href{https://doi.org/10.1103/PhysRevD.104.035011}{\emph{Phys. Rev. D} {\bfseries 104} (2021) 035011} [\href{https://arxiv.org/abs/2104.13523}{{\ttfamily 2104.13523}}].

\bibitem{Okada:2021qmi}
N.~Okada, D.~Raut, Q.~Shafi and A.~Thapa, \emph{{Pseudo-Goldstone dark matter in SO(10)}}, \href{https://doi.org/10.1103/PhysRevD.104.095002}{\emph{Phys. Rev. D} {\bfseries 104} (2021) 095002} [\href{https://arxiv.org/abs/2105.03419}{{\ttfamily 2105.03419}}].

\bibitem{Leblond:2009fq}
L.~Leblond, B.~Shlaer and X.~Siemens, \emph{{Gravitational Waves from Broken Cosmic Strings: The Bursts and the Beads}}, \href{https://doi.org/10.1103/PhysRevD.79.123519}{\emph{Phys. Rev. D} {\bfseries 79} (2009) 123519} [\href{https://arxiv.org/abs/0903.4686}{{\ttfamily 0903.4686}}].

\bibitem{Olmez:2010bi}
S.~Olmez, V.~Mandic and X.~Siemens, \emph{{Gravitational-Wave Stochastic Background from Kinks and Cusps on Cosmic Strings}}, \href{https://doi.org/10.1103/PhysRevD.81.104028}{\emph{Phys. Rev. D} {\bfseries 81} (2010) 104028} [\href{https://arxiv.org/abs/1004.0890}{{\ttfamily 1004.0890}}].

\bibitem{Damour:2001bk}
T.~Damour and A.~Vilenkin, \emph{{Gravitational wave bursts from cusps and kinks on cosmic strings}}, \href{https://doi.org/10.1103/PhysRevD.64.064008}{\emph{Phys. Rev. D} {\bfseries 64} (2001) 064008} [\href{https://arxiv.org/abs/gr-qc/0104026}{{\ttfamily gr-qc/0104026}}].

\bibitem{Hellings:1983fr}
{{Hellings}, R.~W. and {Downs}, G.~S.}, \emph{{Upper limits on the isotropic gravitational radiation background from pulsar timing analysis}}, \href{https://doi.org/10.1086/183954}{\emph{Astrophys. J. Lett.} {\bfseries 265} (1983) L39}.

\bibitem{Mangano:2011ar}
G.~Mangano and P.D.~Serpico, \emph{{A robust upper limit on $N_{\rm eff}$ from BBN, circa 2011}}, \href{https://doi.org/10.1016/j.physletb.2011.05.075}{\emph{Phys. Lett. B} {\bfseries 701} (2011) 296} [\href{https://arxiv.org/abs/1103.1261}{{\ttfamily 1103.1261}}].

\bibitem{LIGOScientific:2021nrg}
{\scshape LIGO Scientific, Virgo, KAGRA} collaboration, \emph{{Constraints on Cosmic Strings Using Data from the Third Advanced LIGO\textendash{}Virgo Observing Run}}, \href{https://doi.org/10.1103/PhysRevLett.126.241102}{\emph{Phys. Rev. Lett.} {\bfseries 126} (2021) 241102} [\href{https://arxiv.org/abs/2101.12248}{{\ttfamily 2101.12248}}].

\bibitem{KAGRA:2013rdx}
{\scshape KAGRA, LIGO Scientific, Virgo, VIRGO} collaboration, \emph{{Prospects for observing and localizing gravitational-wave transients with Advanced LIGO, Advanced Virgo and KAGRA}}, \href{https://doi.org/10.1007/s41114-020-00026-9}{\emph{Living Rev. Rel.} {\bfseries 21} (2018) 3} [\href{https://arxiv.org/abs/1304.0670}{{\ttfamily 1304.0670}}].

\bibitem{Regimbau:2016ike}
T.~Regimbau, M.~Evans, N.~Christensen, E.~Katsavounidis, B.~Sathyaprakash and S.~Vitale, \emph{{Digging deeper: Observing primordial gravitational waves below the binary black hole produced stochastic background}}, \href{https://doi.org/10.1103/PhysRevLett.118.151105}{\emph{Phys. Rev. Lett.} {\bfseries 118} (2017) 151105} [\href{https://arxiv.org/abs/1611.08943}{{\ttfamily 1611.08943}}].

\bibitem{Mentasti:2020yyd}
G.~Mentasti and M.~Peloso, \emph{{ET sensitivity to the anisotropic Stochastic Gravitational Wave Background}}, \href{https://doi.org/10.1088/1475-7516/2021/03/080}{\emph{JCAP} {\bfseries 03} (2021) 080} [\href{https://arxiv.org/abs/2010.00486}{{\ttfamily 2010.00486}}].

\bibitem{Sato_2017}
S.~Sato et~al., \emph{The status of {DECIGO}}, \href{https://doi.org/10.1088/1742-6596/840/1/012010}{\emph{Journal of Physics: Conference Series} {\bfseries 840} (2017) 012010}.

\bibitem{Crowder:2005nr}
J.~Crowder and N.J.~Cornish, \emph{{Beyond LISA: Exploring future gravitational wave missions}}, \href{https://doi.org/10.1103/PhysRevD.72.083005}{\emph{Phys. Rev. D} {\bfseries 72} (2005) 083005} [\href{https://arxiv.org/abs/gr-qc/0506015}{{\ttfamily gr-qc/0506015}}].

\bibitem{Corbin:2005ny}
V.~Corbin and N.J.~Cornish, \emph{{Detecting the cosmic gravitational wave background with the big bang observer}}, \href{https://doi.org/10.1088/0264-9381/23/7/014}{\emph{Class. Quant. Grav.} {\bfseries 23} (2006) 2435} [\href{https://arxiv.org/abs/gr-qc/0512039}{{\ttfamily gr-qc/0512039}}].

\bibitem{Bartolo:2016ami}
N.~Bartolo et~al., \emph{{Science with the space-based interferometer LISA. IV: Probing inflation with gravitational waves}}, \href{https://doi.org/10.1088/1475-7516/2016/12/026}{\emph{JCAP} {\bfseries 12} (2016) 026} [\href{https://arxiv.org/abs/1610.06481}{{\ttfamily 1610.06481}}].

\bibitem{amaroseoane2017laser}
P.~Amaro-Seoane et~al., \emph{Laser interferometer space antenna},  \href{https://arxiv.org/abs/1702.00786}{{\ttfamily 1702.00786}}.

\bibitem{5136190}
P.E.~{Dewdney}, P.J.~{Hall}, R.T.~{Schilizzi} and T.J.L.W.~{Lazio}, \emph{The square kilometre array}, \href{https://doi.org/10.1109/JPROC.2009.2021005}{\emph{Proceedings of the IEEE} {\bfseries 97} (2009) 1482}.

\bibitem{Janssen:2014dka}
G.~Janssen et~al., \emph{{Gravitational wave astronomy with the SKA}}, \href{https://doi.org/10.22323/1.215.0037}{\emph{PoS} {\bfseries AASKA14} (2015) 037} [\href{https://arxiv.org/abs/1501.00127}{{\ttfamily 1501.00127}}].

\bibitem{Thrane:2013oya}
E.~Thrane and J.D.~Romano, \emph{{Sensitivity curves for searches for gravitational-wave backgrounds}}, \href{https://doi.org/10.1103/PhysRevD.88.124032}{\emph{Phys. Rev. D} {\bfseries 88} (2013) 124032} [\href{https://arxiv.org/abs/1310.5300}{{\ttfamily 1310.5300}}].

\bibitem{Schmitz:2020syl}
K.~Schmitz, \emph{{New Sensitivity Curves for Gravitational-Wave Signals from Cosmological Phase Transitions}}, \href{https://doi.org/10.1007/JHEP01(2021)097}{\emph{JHEP} {\bfseries 01} (2021) 097} [\href{https://arxiv.org/abs/2002.04615}{{\ttfamily 2002.04615}}].

\bibitem{Vachaspati:1984gt}
T.~Vachaspati and A.~Vilenkin, \emph{{Gravitational Radiation from Cosmic Strings}}, \href{https://doi.org/10.1103/PhysRevD.31.3052}{\emph{Phys. Rev. D} {\bfseries 31} (1985) 3052}.

\bibitem{Vilenkin:2000jqa}
A.~Vilenkin and E.P.S.~Shellard, \emph{{Cosmic Strings and Other Topological Defects}}, Cambridge University Press (7, 2000).

\bibitem{Vanchurin:2005pa}
V.~Vanchurin, K.D.~Olum and A.~Vilenkin, \emph{{Scaling of cosmic string loops}}, \href{https://doi.org/10.1103/PhysRevD.74.063527}{\emph{Phys. Rev. D} {\bfseries 74} (2006) 063527} [\href{https://arxiv.org/abs/gr-qc/0511159}{{\ttfamily gr-qc/0511159}}].

\bibitem{Ringeval:2005kr}
C.~Ringeval, M.~Sakellariadou and F.~Bouchet, \emph{{Cosmological evolution of cosmic string loops}}, \href{https://doi.org/10.1088/1475-7516/2007/02/023}{\emph{JCAP} {\bfseries 02} (2007) 023} [\href{https://arxiv.org/abs/astro-ph/0511646}{{\ttfamily astro-ph/0511646}}].

\bibitem{Olum:2006ix}
K.D.~Olum and V.~Vanchurin, \emph{{Cosmic string loops in the expanding Universe}}, \href{https://doi.org/10.1103/PhysRevD.75.063521}{\emph{Phys. Rev. D} {\bfseries 75} (2007) 063521} [\href{https://arxiv.org/abs/astro-ph/0610419}{{\ttfamily astro-ph/0610419}}].

\bibitem{Blanco-Pillado:2013qja}
J.J.~Blanco-Pillado, K.D.~Olum and B.~Shlaer, \emph{{The number of cosmic string loops}}, \href{https://doi.org/10.1103/PhysRevD.89.023512}{\emph{Phys. Rev. D} {\bfseries 89} (2014) 023512} [\href{https://arxiv.org/abs/1309.6637}{{\ttfamily 1309.6637}}].

\bibitem{Blanco-Pillado:2017oxo}
J.J.~Blanco-Pillado and K.D.~Olum, \emph{{Stochastic gravitational wave background from smoothed cosmic string loops}}, \href{https://doi.org/10.1103/PhysRevD.96.104046}{\emph{Phys. Rev. D} {\bfseries 96} (2017) 104046} [\href{https://arxiv.org/abs/1709.02693}{{\ttfamily 1709.02693}}].

\bibitem{Cui:2018rwi}
Y.~Cui, M.~Lewicki, D.E.~Morrissey and J.D.~Wells, \emph{{Probing the pre-BBN universe with gravitational waves from cosmic strings}}, \href{https://doi.org/10.1007/JHEP01(2019)081}{\emph{JHEP} {\bfseries 01} (2019) 081} [\href{https://arxiv.org/abs/1808.08968}{{\ttfamily 1808.08968}}].

\bibitem{Buchmuller:2021mbb}
W.~Buchmuller, V.~Domcke and K.~Schmitz, \emph{{Stochastic gravitational-wave background from metastable cosmic strings}}, \href{https://doi.org/10.1088/1475-7516/2021/12/006}{\emph{JCAP} {\bfseries 12} (2021) 006} [\href{https://arxiv.org/abs/2107.04578}{{\ttfamily 2107.04578}}].

\bibitem{Auclair:2019wcv}
P.~Auclair et~al., \emph{{Probing the gravitational wave background from cosmic strings with LISA}}, \href{https://doi.org/10.1088/1475-7516/2020/04/034}{\emph{JCAP} {\bfseries 04} (2020) 034} [\href{https://arxiv.org/abs/1909.00819}{{\ttfamily 1909.00819}}].

\bibitem{Cui:2019kkd}
Y.~Cui, M.~Lewicki and D.E.~Morrissey, \emph{{Gravitational Wave Bursts as Harbingers of Cosmic Strings Diluted by Inflation}}, \href{https://doi.org/10.1103/PhysRevLett.125.211302}{\emph{Phys. Rev. Lett.} {\bfseries 125} (2020) 211302} [\href{https://arxiv.org/abs/1912.08832}{{\ttfamily 1912.08832}}].

\bibitem{Lazarides:2023ksx}
G.~Lazarides, R.~Maji and Q.~Shafi, \emph{{Superheavy quasistable strings and walls bounded by strings in the light of NANOGrav 15~year data}}, \href{https://doi.org/10.1103/PhysRevD.108.095041}{\emph{Phys. Rev. D} {\bfseries 108} (2023) 095041} [\href{https://arxiv.org/abs/2306.17788}{{\ttfamily 2306.17788}}].

\end{thebibliography}\endgroup

\end{document}